\documentclass[11 pt]{article}
\usepackage{epsfig}
\begin{document}

\title{Supermassive screwed cosmic string in dilaton gravity }
\author{V. B. Bezerra\footnote{valdir@fisica.ufpb.br}\\
Departamento de F\'{\i}sica, Universidade Federal da Para\'{\i}ba, \\
58059-970, Jo\~ao Pessoa, PB, Brazil,\\ \\
Cristine N. Ferreira \footnote{crisnfer@cefetcampos.br}\\
N\'ucleo de F\'{\i}sica, Centro Federal de Educa\c c\~ao Tecnol\'ogica de
Campos, \\ Rua Dr. Siqueira, 273 - Parque Dom Bosco, \\28030-130, Campos
dos Goytacazes, RJ, Brazil,\\ \\
H. J. Mosquera Cuesta\footnote{hermanjc@cbpf.br}\\  Instituto de Cosmologia,
Relatividade e Astrof\'{\i}sica (ICRA-BR) \\
Centro Brasileiro de Pesquisas F\'{\i}sicas, Rua Dr. Xavier Sigaud
150, Urca \\ 22290-180, Rio de Janeiro, RJ, Brazil,}

\date{\today}

\maketitle

\begin{abstract}
The early Universe might have undergone phase transitions at energy scales
much higher than the one corresponding to the Grand Unified Theories (GUT)
scales.  At these higher energy scales, the transition at which gravity
separated from all other interactions; the so-called Planck era, more massive
strings called supermassive cosmic strings, could have been produced,
with energy of about $10^{19}GeV$. The dynamics of strings formed with
this energy scale cannot be described by means of the weak-field
approximation, as in the standard procedure for ordinary GUT cosmic
strings. As suggested by string theories, at this extreme energies,
gravity may be transmitted by some kind of scalar field (usually called
the {\it dilaton}) in addition to the tensor field of Einstein's theory
of gravity.  It is then permissible to tackle the issue regarding
the dynamics of supermassive cosmic strings within this framework. With
this aim we obtain the gravitational field of a supermassive screwed
cosmic string in a scalar-tensor theory of gravity.  We show that for
the supermassive configuration exact solutions of scalar-tensor screwed
cosmic strings can be found in connection with the Bogomol'nyi limit. We
show that the generalization of Bogomol'nyi arguments to the Brans-Dicke
theory is possible when torsion is present and we obtain an exact solution
in this supermassive regime, with the dilaton solution obtained by consistency
with internal constraints.
\end{abstract}

\newpage

\section{Introduction}

Topological defects such as cosmic strings\cite{Kibble,Vilenkin} are predicted
to form in the early Universe, at GUT scale, as a result of symmetry-breaking
phase transitions envisaged in gauge theories of elementary particles interactions.
These topological defects may also help to explain the most energetic events in
the Universe, as the cosmological Gamma-Ray Bursts (GRBs)\cite{brandenb93}, very
high energy neutrinos\cite{Brandenberger} and gravitational-wave bursts\cite{HD00}.

Cosmic strings can also be formed by phase  transitions at energy scales
higher than the GUT scale, which result in the production of more massive
strings. The current theory tells us that these strings had been produced
before or during inflation, so that their dynamical effects would not leave any
imprints in the Cosmic Microwave Background radiation(CMB). These cosmic
strings are referred to as supermassive cosmic strings and had an energy of
approximately three orders of magnitude higher than the ordinary cosmic string.
This means that they can no longer be treated by using the weak-field
approximation.

The space-time of supermassive cosmic strings has been examined by Laguna and
Garfinkle\cite{Laguna} following a method developed by Gott\cite{Gott}. Their
approach considered carefully the possible asymptotic behavior of the supermassive
cosmic string metric. Their conclusions were that such a string has a Kasner-type
metric outside the core with an internal metric that is regular.

 Some authors have studied solutions corresponding to topological defects in
different contexts like in Brans-Dicke \cite{rom}, dilaton theory \cite{Ruth},
and in more general scalar-tensor couplings \cite{mexg,Cris1}.  In this paper
we study, within the framework of a scalar-tensor theory of gravity, the dynamics
of matter in the presence of a supermassive screwed cosmic string (SMSCS) that had
been produced in the very early Universe.

Scalar-tensor theories of gravity can be considered\cite{Jordan} as the most
promising alternatives for the generalization of Einstein's gravity, and is
motivated by string theory. In scalar-tensor theories, gravity is mediated by
a long-range scalar field in addition to the usual tensor field present in
Einstein's theory \cite{Jordan}. Scalar-tensor theories of gravity are currently
of particular interest since such theories appear as the low energy limit of
supergravity theories constructed from string theories\cite{Green} and other
higher dimensional gravity theories\cite{Applequist}. However, due to the lack
of a full non-perturbative formulation, which allows a description of the early
Universe close to the Planck time, it is necessary to study classical cosmology
prior to the GUT epoch by recurring to the low-energy effective action induced
by string theory. The implications of such actions for the process of structure
formation have been studied recently\cite{Cris1,Emilia1}.

This work is outlined as follows. In Section 2, we describe the configuration of
a supermassive screwed cosmic string (SMSCS) in scalar-tensor theory of gravity.
In Section 3, we construct the Bogomol'nyi conditions of a screwed cosmic string
in scalar-tensor theory. In Section 4, we solve the exact equations for the
exterior spacetime, by applying Linet's method\cite{Linet}. Then, we match the
exterior solution to the internal one. We also derive the deficit angle associated
with the metric of a SMSCS. Section 5 discusses the particle motion around this
kind of defect. In Section 6, our discussion and conclusions are presented.

\section{Cosmic string in dilaton-torsion gravity }

The scalar-tensor theory of gravity with torsion can be described by the action
which is given (in the Jordan-Fierz frame) by

\begin{equation}
{\cal S} = \frac{1}{16\pi} \int d^4x \sqrt{-\tilde{g}}
\left[\tilde{\Phi}\tilde{R} - \frac{\omega(\tilde{\Phi})}{\tilde{\Phi}}
\tilde{g}^{\mu\nu} \partial_{\mu}\tilde{\Phi}\partial_{\nu}\tilde{\Phi}
\right] + {\cal S}_{m}[\Psi_m, \tilde{g}_{\mu\nu}] , \label{1}
\end{equation}

\noindent
where ${\cal S}_m$ is the action of all matter fields and $\tilde{g}_{\mu\nu}$
is the metric in the Jordan-Fierz frame. In this frame, the Riemann curvature
scalar, $\tilde{R}(\{\})$, can be written as

\begin{equation}
\tilde R = \tilde R(\{\}) +
\epsilon \frac{\partial_{\mu} \tilde \Phi \partial^{\mu} \tilde \Phi}{\tilde \Phi^2}  \; ,
\end{equation}

\noindent
with $\epsilon $ being the torsion coupling constant \cite{Cris1}. Note that in
this theory matter couples minimally and universally with
$\tilde{g}_{\mu\nu}$ and not with $\tilde{\Phi}$.

For technical reasons, working in the Einstein or conformal frame is more simple.
In this frame, the kinematic terms associated with the scalar and tensor fields
do not mix and the action is given by

\begin{equation}
{\cal S} = \frac{1}{16\pi G} \int d^4x \sqrt{-g} \left[ R(\{\}) -
2\kappa(\phi)g^{\mu\nu}\partial_{\mu}\phi\partial_{\nu}\phi \right]
+ {\cal S}_{m}[\Psi_m,\Omega^2(\phi)g_{\mu\nu}],\label{8}
\end{equation}

\noindent
where $g_{\mu\nu}$ is a pure rank-2 tensor in the Einstein frame and $\Omega
(\phi)$ is an arbitrary function of the scalar field.

Action (\ref{8}) is obtained from (\ref{1}) by a conformal transformation

\begin{equation}
\tilde{g}_{\mu\nu} = \Omega^2(\phi)g_{\mu\nu} ,\label{9}
\end{equation}

\noindent
and by a redefinition of the quantity

\begin{equation}
G\Omega^2(\phi) = \tilde{\Phi}^{-1} \; ,
\label{vbb1}
\end{equation}
\noindent
which makes evident that any gravitational phenomena will be affected by
the variation of the gravitation ``constant" $G$ in scalar-tensor theories
of gravity. Let us introduce a new parameter, $\alpha $, such that

\begin{equation}
\alpha^2 \equiv \left( \frac{\partial \ln \Omega(\phi)}{\partial
\phi} \right)^2 = [2\omega(\tilde{\Phi}) + 3]^{-1} ,
\end{equation}
\noindent
which can be interpreted as the (field-dependent) coupling strength between
matter and the scalar field.  In order to make our calculations as general as
possible, we will not specify the factors $\Omega(\phi)$ and $\alpha (\phi)$,
thus leaving them as arbitrary functions of the scalar field.

In the conformal frame, the Einstein equations read as

\begin{equation}
G_{\mu\nu}  = 2\kappa \partial_{\mu}\phi\partial_{\nu}\phi -
\kappa g_{\mu\nu}g^{\alpha\beta}\partial_{\alpha}\phi\partial_{\beta}
\phi + 8\pi G T_{\mu\nu}\; .\label{10}
\end{equation}

\noindent
where $\kappa $ is a function  defined by

\begin{equation}
\kappa = 1 -2\epsilon \alpha(\phi)^2 \; ,\label{kappa}
\end{equation}

\noindent
which explicitly depends on the scalar field and torsion. The energy-momentum
tensor is defined as

\begin{equation}
T_{\mu\nu} \equiv \frac{2}{\sqrt{-g}}\frac{\delta {\cal S}_m}
{\delta g_{\mu\nu}},
\label{11}
\end{equation}

\noindent
which is no longer conserved. It is clear from (\ref{9}) that we can relate
the energy-momentum tensor in both frames in such a way that

\begin{equation}
\tilde{T}^{\mu}_{\nu} = \Omega^{-4}(\phi)T^{\mu}_{\nu}\; .\label{Ttt}
\end{equation}

For the sake of simplicity, in what follows we will focus on the Brans-Dicke
theory where $\Omega^2(\phi) = e^{2\alpha \phi}$. To account for the (unknown)
coupling of the cosmic string field with the dilaton, we will choose the action,
$ \cal S $, in the Einstein frame, as

\begin{equation}
{\cal S} = \frac{1}{16\pi G} \int d^4x \sqrt{-g}
\left[R - 2g^{\mu\nu} \kappa \partial_{\mu}\phi\partial_{\nu}\phi +
e^{2(a + 2)\alpha \phi}{\cal L}_{m}[\psi_m,e^{2 \alpha \phi}
g_{\mu\nu}] \right]\; , \label{1c}
\end{equation}

\noindent
where $a$ is an arbitrary parameter which couples the dilaton to the matter
fields.

Taking into consideration the symmetry of the source, we impose that the metric
is static and cylindrically symmetric. Thus, let us choose a general
cylindrically symmetric metric with coordinates $(t, \rho, z, \theta)$ as

\begin{equation}
ds^2 = e^{\gamma }(-dt^2 + d\rho^2 + dz^2) + \beta^2 e^{-\gamma}
d\theta^2,\label{12}
\end{equation}

\noindent
where the metric functions $\gamma$ and $\beta$ are functions of the radial
coordinate $\rho$ only. In addition,  the metric functions satisfy the regularity
conditions at the axis of symmetry ($\rho=0$)

\begin{equation}
\gamma = 0,  \;\; \frac{d\gamma}{d\rho}=0 \;\;
\mbox{and} \;\; \frac{d\beta}{d\rho} =1 .\label{13}
\end{equation}

Next we will search for a regular solution of a self-gravitating vortex in
the framework of a scalar-tensor gravity. Hence, the simplest bosonic vortex
arises from the Lagrangean of the Abelian-Higgs $U(1) $ model, which contains
a complex scalar field plus a gauge field and can be written as

\begin{equation}
{\cal L}_m  = -\frac{1}{2}\tilde{g}^{\mu\nu}D_{\mu}\varphi D_{\nu}\varphi^*
- \frac{1}{4} \tilde{g}^{\mu\nu}\tilde{g}^{\alpha\beta}F_{\mu\alpha}
F_{\nu\beta} - V(\mid\varphi\mid ) \label{2}
\end{equation}

\noindent
where $D_{\mu}\varphi \equiv (\partial_{\mu} + ieA_{\mu})\varphi$ and
$F_{\mu\nu}$ is the field-strength associated with the vortex $A_{\mu}$.
The potential is ''Higgs-inspired'' and contains appropriate $\varphi$
interactions so that there occurs a spontaneous symmetry breaking. It is
given by

\begin{equation}
V(\mid\varphi\mid ) = \frac{\lambda_{\varphi}}{4}
(\mid\varphi\mid^2 - \eta^2)^2 ,\label{3}
\end{equation}

\noindent
with $\eta $ and $\lambda_{\phi}$ being positive parameters. A vortex
configuration arises when the $U(1)$ symmetry associated with the $(\varphi,
A_{\mu})$ pair is spontaneously broken in the core of the vortex. In what
follows, we will restrict ourselves to configurations corresponding to an
isolated and static vortex along the $z$-axis. In a cylindrical coordinate
system $(t,\rho,\theta,z)$, such that $\rho \geq 0$ and $0 \leq \theta <2\pi$,
we make the choice

\begin{equation}
\varphi = R(\rho)e^{i\theta} \;\; \mbox{and} \;\;  A_{\mu} = \frac{1}{q}[P(\rho) - 1]
\delta^{\theta}_{\mu} , \label{4}
\end{equation}

\noindent
in analogy with the case of ordinary (non-conducting) cosmic strings. The
variables $R, P$ are functions of $\rho$ only. We also require that these
functions are regular everywhere and {\bf so} they must satisfy the usual
boundary conditions for a vortex configuration\cite{niel}, which are the
following

\begin{eqnarray}
R(0) = 0 \;\; \mbox{and} \;\; P(0) =1 \nonumber\\
\lim_{\rho \rightarrow\infty} R(\rho)=\eta \;\; \mbox{and}
\lim_{\rho \rightarrow \infty} P(\rho) = 0 . \label{5}
\end{eqnarray}

With the metric given by Eq.(\ref{12}) we are in a position to write down
the full equations of motion for the self-gravitating vortex in a scalar-tensor
gravity. In the conformal frame these equations reduce to

\begin{eqnarray}
\beta^{''} & = & 8\pi G\beta  [T^t_t + T^{\rho}_{\rho}] e^{\gamma} \label{derivbeta}\\
(\beta \gamma')' &= &8 \pi G  \beta [ T^{\rho}_{\rho} +
T^{\theta}_{\theta}] e^{\gamma },\label{derivbeta2}\\
\beta^{'}\gamma^{'} & = & \frac{\beta(\gamma^{'})^{2}}{4} -  \kappa \beta(\phi^{'})^{2} +
8\pi G e^{\gamma }T^{\rho}_{\rho} \label{gamma1} \\
(\kappa \beta\phi^{'})^{'} & = & - 4\pi G \alpha \beta \left[ (a+1) T^t_t + T^{\rho}_{\rho} +
T^{\theta}_{\theta}\right] e^{\gamma } ,\label{14}
\end{eqnarray}

\noindent
where $(')$ denotes derivative with respect to $\rho$.

 In what follows we will analyze the Bogomonl'nyi conditions for this
supermassive cosmic string in this dilaton-torsion theory. In some models
 \cite{gibbons88,yoonbai94},
it is possible to work with first order differential equations, which are
called Bogomol'nyi equations, instead of more complicated Euler-Lagrange
equations. This approach can be applied to Eqs.(\ref{derivbeta}-\ref{14})
when $\kappa =0 $ and $ a=-1 $. The traditional method to obtain such
equations \cite{gibbons88,yoonbai94} is based on rewriting an expression
for the energy of a field configuration, in such a way that it has a lower
bound, which is of topological origin. The field configurations which
saturate this bound satisfy Euler-Lagrange equations as well as Bogomol'nyi
equations.

\section{Bogomol'nyi bounds for dilatonic supermassive cosmic strings }

The gravitational field produced by a supermassive cosmic string in the
framework of general relativity has been studied by numerical method
\cite{Laguna} in the case $G\mu = 1/4$, where $\mu$ is the linear mass
density of the string, in the Bogomonl'nyi limit. In this case, the
exact field equations lead to the metric of a cylindrical spacetime
 \cite{yoonbai94,Linet1,GLL}. In this section, we analyse the
possibility of finding a Bogomol$'$nyi bound for a dilatonic-torsion
cosmic string. It is worth commenting that in the usual Brans-Dicke
theory this limit is not possible\cite{Ruth}, but in the space-time
with torsion it is possible.  In this case the dilaton does not have any
dynamics, that is, $\kappa =0$, but even in this situation the solution
is non-trivial. As the Bogomonl'nyi equations are differential equations
of first order, as a consequence in the supermassive screwed cosmic string
case we can find an exact solution. In order to do this we have to compute
the energy-momentum tensor which can be written as

\begin{equation}
T^{\mu}_{\nu} = 2 g^{\mu \alpha}\frac{\partial {\cal L}}{\partial
g^{\alpha \nu}} - \delta^{\mu}_{\nu}  {\cal L},
\end{equation}

\noindent
whose non-vanishing components are

\begin{equation}
T^{t}_{t}  =  T^z_z =   - \frac{1}{2} e^{2(a+1)\alpha \phi} \{
e^{-\gamma}R'^{2} + \frac{e^{\gamma }}
{\beta^2}R^2P^2   + \frac{e^{-2\alpha \phi }}{\beta^2}
(\frac{P'^2}{4\pi q^2}) + 2e^{2\alpha \phi}V(R) \}\label{Tt}\; ,
\end{equation}

\noindent
Note that from (\ref{Tt}) we see that the boost symmetry is not preserved.
This fact is related with the absence of current in $z$-direction, as in the
usual ordinary cosmic string.

The transverse components of the energy-momentum tensor are given by the
expressions

\begin{equation}
T^{\rho}_{\rho} =   \frac{1}{2}e^{2(a+1)\alpha \phi} \{ e^{- \gamma}
R'^2 - \frac{e^{\gamma}}{\beta^2} R^2P^2 +\frac{e^{-2\alpha \phi }}
{\beta^2}(\frac{P'^2}{4\pi q^2}) - 2e^{2\alpha \phi}
V(R) \} \label{Tr}\; ,
\end{equation}

\begin{equation}
T^{\theta}_{\theta}  =   - \frac{1}{2}e^{2(a+1)\alpha \phi}
\{ e^{- \gamma}R'^2  - \frac{e^{\gamma }}{\beta^2} R^2P^2
- \frac{e^{-2\alpha \phi }}{\beta^2}(\frac{P'^2}{4\pi q^2})
+ 2e^{2\alpha \phi} (\phi)V(R ) \}  \label{Ttheta} \; .
\end{equation}

As stated earlier, the energy-momentum tensor is not conserved in the
conformal frame. Thus, in this situation, we have

\[\nabla_{\mu}T^{\mu}_{\nu} = \alpha(\phi) T \nabla_{\nu}\phi ,\]

\noindent
where $T$ is the trace of the energy-momentum tensor. This equation gives an
additional relation between the scalar field $\phi$ and the source described
by $T^\mu_\nu$. This energy-momentum tensor has only an $\rho$-dependence. The
matter field equations can be written as

\begin{equation}
R'' + R'\left[\frac{\beta'}{\beta} -2(a+1)\alpha \phi' \right] -
R\left[\frac{P^2}{\beta^2} + 4\lambda e^{2\alpha \phi}(R^2-\eta^2)\right] = 0\; ,
\end{equation}

\begin{equation}
P'' + P'\left[\frac{\beta'}{\beta} -2(a+2)\alpha \phi' \right] - q^2R^2P
e^{2\alpha \phi} =0 \; .
\end{equation}

Now let us define the energy density per unit of length of the string (in
Jordan-Fierz frame), $\mu$, as

\begin{equation}
\mu = 2\pi \int_0^{\infty}\sqrt{-\tilde g} \tilde T_t^t d\rho \; \label{mass}.
\end{equation}

Considering the $t$-component of the energy-momentum tensor given by (\ref{Tt})
we realize that $a=-1 $ is a rather special point in the analysis of the
Bogomol'nyi solution. Following the studies in the literature \cite{Ruth,Linet}
we analyzed the case where $\gamma =0$. In such a situation it is possible
to find the Bogomonl'nyi solution for $\kappa =0$ because this is the only value
of this parameter that turns equation (\ref{gamma1}) consistent with the fact
that the transverse component of energy-momentum tensor vanishes. Considering
the Einstein frame and the expression for $T^t_t$ given by Eq.(\ref{Tt}),
Eq.(\ref{mass}) turns into

\begin{equation}
\begin{array}{ll}
\mu =& 2 \pi  \int_0^{\infty} \beta d \rho\left[ \frac{1}{2}
\left(\frac{P'}{q \beta} e^{-\alpha \phi}
+ e^{\alpha \phi} \sqrt{2\lambda }( R^2 - \eta^2)\right)^2 +
\frac{1}{q \beta }\sqrt{2\lambda }(R^2 - \eta^2) P' + \right. \\
& \left.+ \left(R' + \frac{RP}{\beta}\right)^2 -
\frac{2}{\beta}e^{2\alpha \phi}RPR' \right],
\end{array}\label{bog11}
\end{equation}

where we have taken into account relations between $\tilde T_{\nu}^{\mu}$ and
$T_{\nu}^{\mu}$ and$\tilde g_{\mu \nu}$ and $g_{\mu \nu}$, given by Eqs. (\ref{9})
and (\ref{Ttt}), respectively. Using the Bogomonl'nyi method we have that the
quadratic terms in (\ref{bog11}) vanish and therefore we get

\begin{equation}
\mu = 2 \pi  \int_0^{\infty} \beta d \rho \left[
\frac{1}{q \beta }\sqrt{2\lambda }(R^2 - \eta^2) P' -
\frac{2}{\beta}e^{2\alpha \phi}RPR' \right].
\end{equation}

Let us analyze the behavior of the dilaton field and the supermassive cosmic
string limit. To do this, let us write

\begin{equation}
\mu = \pi \int_0^{\infty}  [(R^2 - \eta^2)P]' \; d \rho
\end{equation}

\noindent
where the Bogomol'nyi limit, $8\lambda = q^2$, was taken into account.
Integrating by parts and using the regularity conditions (\ref{13}) and
the boundary conditions (\ref{5}), we find

\begin{equation}
\mu \geq \pi\eta^2 \; ,
\end{equation}

\noindent
with $\mu = \pi \eta^2$ being the lower bound which corresponds to a solution
in the Bogomol$'$nyi limit. We can see that there is no force between vortices
and also that the equations of motion are of first order. The search of the
Bogomol$'$nyi bound for the energy, in the Bogomol$'$nyi limit, yields the
following system of equations

\begin{equation}
R'= -\frac{RP }{\beta }\label{bog1}\; ,
\end{equation}

\begin{equation}
P'= -\frac{q^2}{2} \beta e^{2\alpha \phi} (R^2 - \eta^2)\label{bog2}\; .
\end{equation}

>From these equations one can see that the transversal components of the
energy-momentum tensor density vanish, that is, $T^{\rho}_{\rho} =
T^{\theta}_{\theta} =0$, while the non-zero components $T^t_t$ and $T^z_z$,
are related by $T^t_t = T^z_z$. This corresponds to the minimal configuration
asociated with a cosmic string in scalar-tensor theories of gravity.

In the ordinary case, with no torsion, this configuration is not allowed
because $\epsilon =0 $ and as a consequence there is no possibility to obtain
the Bogomonl'nyi configuration with $\phi \neq 0$. Then, the presence of the
torsion is of fundamental importance to obtain the Bogomonl'nyi bound in the
framework of scalar-tensor theories.

With these conditions one can simplify Eq.(\ref{derivbeta}), which results in

\begin{equation}
\beta'' = -8\pi G [ \frac{R^2P^2}{\beta} + \frac{q^2}{4}
\beta (R^2 -\eta^2)^2 e^{2\alpha \phi}]\label{beta} \; .
\end{equation}

Using Eqs.(\ref{bog1}), (\ref{bog2}) in (\ref{beta}) we find

\begin{equation}
\beta'' = 8\pi G \left[ RPR' + \frac{1}{2} (R^2 -\eta^2)P' \right]
\label{bog3}\; ,
\end{equation}
which has the usual form obtained in the context of general relativity. One
can also see that the dilaton only appears in Eq.(\ref{bog2}). Then, it is
possible to evaluate the integral of Eq.(\ref{bog3}), which gives us

\begin{equation}
\beta'= - 4 \pi G (R^2 - \eta^2)P + 1 - 4\pi G \eta^2 \label{bog4} \; .
\end{equation}

In this section we have analyzed the conditions to obtain the Bogomonl'nyi
the bounds for a SMSCS. We note that for $a= - 1$ with $\kappa =0$ it is
possible that this topological bound can be saturated. In this case, we
can see from Eq.(\ref{kappa}) that the scalar-tensor parameter $\alpha $
is connected with the torsion through the relation $\alpha^2 = \frac{1}{2
\epsilon}$. In fact, in the Einstein frame, Eq.(\ref{10}) does not give
us the contribution from the dilaton-torsion term. The modification
introduced in this limit, that could have occurred in the early Universe,
appears in Eq.(\ref{bog2}). The $\phi $-solution can be valuable in connection
with the cosmic string fields. In the next section, we will attempt to solve
the field equations (\ref{bog1}), (\ref{bog2}) and (\ref{bog4}) in the limit
of the supermassive screwed cosmic string. For this purpose we divide the
space-time in two regions:  an exterior region $ \rho > \rho_0$ and an interior
region $\rho \leq \rho_0$, where all the string fields contribute to the
energy-momentum tensor. Once these separate solutions are obtained, we then
match them providing a relationship between the internal parameters of the
string and the space-time geometry outside it.

\section{Supermassive limit of dilatonic cosmic strings }

Now, let us analyze the possibility of obtaining an exact solution of the
first-order differential equations studied in the last section. It is possible
to find the supermassive limit of a $U(1)$ gauge cosmic string in scalar-tensor
theories of gravity with torsion. Using the transformations \cite{Shaver}

\begin{equation}
\begin{array}{ll}
\beta = \frac{2 \sqrt{2}}{\eta q} \bar \beta \; ,\\
R = \eta \bar R \; ,\\
\rho = \frac{2 \sqrt{2}}{\eta q} \bar {\rho} \; ,
\end{array}
\end{equation}

\noindent
and the relation $\bar \eta^2 = 4 \pi G \eta^2 $,  we obtain the equations
of the motion in the supermassive limit $\bar \eta^2 = 1$. They are given by

\begin{equation}
\bar \beta ' = - \bar P (\bar R^2 -1) \;  ,
\end{equation}

\begin{equation}
\bar P ' = 4 \bar \beta (\bar R^2 -1) e^{2\alpha \phi } \; ,
\end{equation}

\begin{equation}
\bar R ' = \frac{\bar R \bar P }{\bar \beta } \; .
\end{equation}

Now, we consider the solutions of these equations in two regions: the internal,
where $\bar R^2 <<1$, and the external one. As stressed earlier, we do not
use the weak field approximation to derive this internal solution because at the
energy scale of the transition which allows the formation of such supermassive
comsic string, such an approach cannot be applied.

\subsection{Internal exact solution}

To solve these equations we make the assumption that $\bar R^2 << 1$, if we
consider that the radius of the string is of the order of $10^{-30} cm $. Thus,
the interior solution of the field equations, in the first order approximation
in $\bar R$, may be taken as

\begin{equation}
\bar \beta ' =  \bar P \; ,
\end{equation}

\begin{equation}
\bar P ' = -4 \bar \beta  e^{2\alpha \phi } \; ,
\end{equation}

\begin{equation}
\bar R ' = \frac{\bar R \bar P }{\bar \beta }\label{R} \; .
\end{equation}

The solution then reads

\begin{equation}
\bar \beta = (1+m)^{-1}\left[m \bar {\rho} + \frac{1}{2} \sin(2 \bar
{\rho})\right]\; ,
\end{equation}

\begin{equation}
\bar P = (1+m)^{-1}[m+\cos(2 \bar {\rho})]\; ,
\end{equation}
where $m$ is a constant.

Now, let us consider the solution for the field $\bar R$ taking into
account Eq.(\ref{R}). Thus, we have

\begin{equation}
\bar  R = B(1+m)^{-1} \left[m \bar{\rho}  + \frac{1}{2} \sin(2 \bar
{\rho})\right]\; ,
\end{equation}

\noindent
with  $B = (1+m)^{-1} \left[ m \bar {\rho}_0 + \frac{1}{2} \sin(2 \bar
{\rho}_0) \right]$, in such a way that at the boundary $\bar {\rho}= \bar
{\rho}_0$, $\bar R(\bar {\rho}_0)=1$. The dilaton solution thus reads

\begin{equation}
\phi = \frac{1}{2\alpha } \ln\left\{\frac{\sin(\sqrt{\frac{\lambda}{\pi}}
{\rho})}{[m \sqrt{\frac{\lambda}{\pi}} {\rho} + \sin(\sqrt{\frac{\lambda}{\pi}}
{\rho})]}\right\} \; ,
\end{equation}

\noindent
with ${\rho} = \sqrt{x^2 + y^2}$ and the $\phi$ solution given in Fig.1.

\begin{figure}
\centering
\mbox{{\epsfig{figure=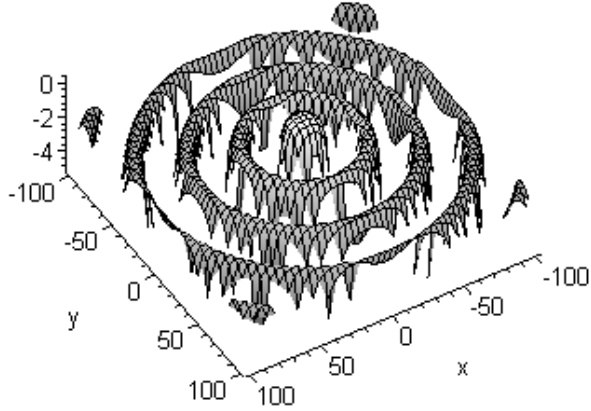,width=0.8\textwidth}}}
\vskip -1 cm
\caption{\label{fig1} The $\phi $ solution as a function of x and y for
$ m= 0.2$,\, \,$\sqrt{\frac{\lambda}{\pi}}=0.2 $ and $\alpha = 0.5$, with
x and y in the range  \, \, $-100 \leq x \leq 100$ and  $-100\leq y \leq 100$.}
\end{figure}

The Fig. \ref{fig1} shows the exact solution for a supermassive screwed cosmic
string when the limit $\bar R<<1$ is applicable.

\subsection{External solution and the matched conditions}

Now we will analyze the assymptotic solution for  $\rho\geq \rho_0$ and apply
the boundary and matching conditions. To do this we have to obtain the exterior
solution. As this solution corresponds to the  exterior region, thus we consider
$T^{\mu }_{\nu} =0$. For this region, the metric in the Einstein frame takes the
form

\begin{equation}
ds^2 = -dt^2 + d{\rho}^2 + dz^2 + \beta_{\infty}^2 d\theta^2\label{ametric}\; ,
\end{equation}

\noindent
in an appropriate coordinate system where the constant $\beta_{\infty}$ fixes
the radius of the compactified dimension. This asymptotic form of the metric
given by Eq.(\ref{ametric}) for a cosmic string has been already examined in
literature \cite{Linet1}. In the case of a space-time with torsion, we can find
the matching conditions using the fact that$[\{^\alpha_{\mu\nu}\}]_{_{{\rho} =
{\rho}_0}}^{(+)} = [\{^\alpha_{\mu\nu}\}]_{_{{\rho}={\rho}_0}}^{(-)}$, and the
metricity constraint $[g_{\mu \nu}K_{(\mu \nu)}^{\hspace{.3 true cm}
\alpha}]_{_{{\rho}={\rho}_0}}^{+}= [g_{\mu \nu}K_{(\mu \nu)}^{\hspace{.3 true cm}
\alpha}]_{_{{\rho}={\rho}_0}}^{-}$, for the contorsion. Thus, we find the
following continuity conditions

\begin{eqnarray}
&[g_{\mu \nu}]_{_{{\rho}={\rho}_0}}^{(-)} =
[g_{\mu \nu}]_{_{{\rho}={\rho}_0}}^{(+)}, \nonumber \\
&[\frac{\partial g_{\mu
\nu}}{\partial x^{\alpha }}]_{_{{\rho}={\rho}_0}}^{(+)}  =
[\frac{\partial g_{\mu \nu}}{\partial x^{\alpha}}]_{_{{\rho}={\rho}_0}}^{(-)},
\label{junc1}
\end{eqnarray}

\noindent
where $(-)$ represents the internal region and $(+)$ corresponds to the
external region around ${\rho} = {\rho}_0$\cite{Kopczynski,Volterra}. For
a supermassive U(1) gauge cosmic string in the Bogomonl'ny limit where $q^2
= 8 \lambda$ and  $\mu = \pi \eta^2 $, thus we find that

\begin{equation}
\frac{2\sqrt{2}}{\eta q}\frac{1}{1+m}\left[m \bar {\rho}_0 + \frac{1}{2}
\sin(2 \bar {\rho}_0)\right] = \beta_{\infty}\; ,
\end{equation}

\noindent
and

\begin{equation}
\frac{1}{1+m}\left[m  +  \cos(2\bar {\rho}_0)\right] = 0 \; .
\end{equation}

\noindent
For $\cos 2\bar {\rho}_{0} = - m$ and $\sin 2\bar {\rho}_{0}= \sqrt{1-m^2}$,
the solution for the long range field $\phi $, in $\bar {\rho} = \bar{\rho}_{0}$,
is given by

\begin{equation}
\phi(\bar{\rho}_0) = \frac{1}{2\alpha}{\bf \ln} \left\{ \frac{\sqrt{1- m^2}}{[2
m \bar{\rho}_0 + \sqrt{1- m^2}]}\right\} \; .
\end{equation}

Therefore, the consistent asymptotic solution for the fields in $ {\rho}\geq
{\rho}_0$ are

\begin{equation}
\begin{array}{ll}
\bar P = 0,\\
\bar R= 1 \;  .
\end{array}
\end{equation}

This external solution is consistent with the boundary condition given by
Eq.(\ref{5}) for $\bar \eta^2 =1$. Thus, the expression for the linear energy
density turns into

\begin{equation}
\mu = \int_0^{{\rho}_0} \int_0^{2\pi} T^t_t
\beta d\theta d {\rho} = \frac{1}{4} \left\{ 1 -
(1+m)^{-1}[m + \cos (2 \bar {\rho}_0)]\right\} = \frac{1}{4} \; .
\end{equation}

The constant $m$ can be restricted to the range $-1<m< 1$. In the Bogomonl'nyi
limit, thus the assymptotic form (\ref{ametric}) has the coefficient
$\beta_{\infty}$ given by

\begin{equation}
\beta_{\infty} = 2\sqrt{\frac{\pi }{\lambda}}\frac{1}{1+m}\left[m
\bar {\rho}_0 + \frac{1}{2} \sqrt{1-m^2}\right]\; .
\end{equation}

\noindent
One can note that when $m =0$, we have that $\beta_{\infty} = \sqrt{\frac{\pi}
{\lambda}}$, and therefore, the contribution arising from the scalar field
vanishes, due to the fact that $\phi =0$. In this case we recover the
Einstein solution for the supermassive cosmic string\cite{Linet1}. Therefore,
the asymptotic metric in the Jordan-Fierz frame is given by

\begin{equation}
ds^2 =
\frac{\sin(\sqrt{\frac{\lambda}{\pi}}{\rho})}{\left[\sqrt{\frac{\lambda}{\pi}}m
{\rho} + \sin(\sqrt{\frac{\lambda}{\pi}}{\rho})\right]} \left\{-dt^2 + d{\rho}^2
+ \beta_{\infty} d\theta^2 + dz^2 \right\} \label{m5} \; .
\end{equation}

Thus, in the torsion-dilaton case appears a Newtonian force in the Bogomonl'nyi
limit of the exact solution. There is no a $\phi $-equation in the supermassive
limit, but the $\phi$ solution is found in connection with the internal
solution of the field equations for the cosmic string. The  continuity in the
boundary gives us the $\phi$-exterior solution. It is of the same form as for
the interior solution.

\section{Particle deflection near a supermassive screwed cosmic string }

Now let us  consider some new physical effects derived from the obtained
solution for a supermassive screwed cosmic string. Thus in order to see
these, we study the geodesic equation in the space-time under consideration,
given by Eq.(\ref{m5}), where $g_{tt}$ reads

\begin{equation}
g_{tt} = \frac{\sin(\sqrt{\frac{\lambda}{\pi}} {\rho})} {\left[\sqrt{\frac{\lambda}
{\pi}} m {\rho} + \sin(\sqrt{\frac{\lambda}{\pi}} {\rho})\right]}\label{htt} \; ,
\end{equation}

\begin{figure}
\centering
\mbox{{\epsfig{figure=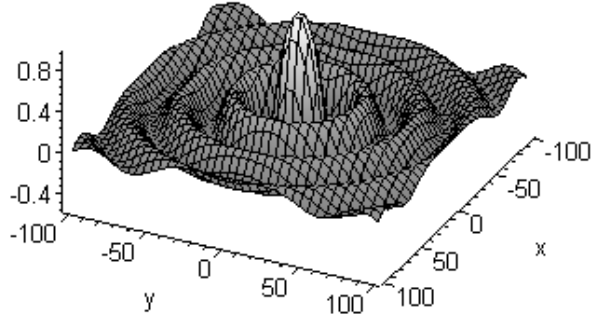,width=.8\textwidth}}}
\vskip -1 cm
\caption{\label{fig2} The tt-component of the metric  solution as a function
of x and y, for $ m= 0.2$,\, \, $\sqrt{\frac{\lambda}{\pi}}=0.2 $ and $\alpha
= 0.5$, with x and y in the range  \, \, $-100 \leq x \leq 100$ and  $-100\leq
y \leq 100$.}
\end{figure}

whose graph is showed in Fig.(\ref{fig2}). We shall consider the effect that
torsion plays on the gravitational force generated by a SMSCS on a particle
moving around it, by assuming that the particle has no charge. To do this,
let us consider the Christoffel symbols in the Jordan-Fierz, $\{^{\mu}_{\alpha
\beta} \}_{JF}$, frame as a sum of the contributions arising from the Christoffel
symbols in the Einstein frame $\{^{\mu}_{\alpha\beta} \}$, from the
contortion given by

\begin{equation}
K_{(\alpha \beta)}^{\hspace{.3 true cm} \mu} =
{\alpha(\phi)\over 2} (\delta^{\mu}_{\alpha}\partial_{\beta} \phi +
\delta^{\mu}_{\beta}\partial_{\alpha} \phi - 2 g_{\alpha \beta} g^{\mu
\nu}\partial_{\nu} \phi) {\bf \; ,}
\end{equation}

\noindent
and from the dilaton field, $\phi$. Thus, summing up these contributions, we
have

\begin{equation}
\{^{\mu}_{\alpha \beta} \}_{JF} = \{^{\mu}_{\alpha \beta} \} +
K_{(\alpha \beta)}^{\hspace{.3 true cm}\mu} + {\alpha(\phi)\over 2}
(\delta^{\mu}_{\alpha}\partial_{\beta} \phi + \delta^{\mu}_{\beta}
\partial_{\alpha} \phi).
\end{equation}

Let us consider the particle moving with speed $|{\bf v}| \equiv
dx^\alpha/dt \leq 1 $, in which case the geodesic equation becomes

\begin{equation}
\frac{d^2x^i}{dt^2} + \{^{i}_{tt} \}_{JF} =0\; ,
\label{geo1}
\end{equation}

\noindent
where  $i$ is a spatial coordinate index. In the Einstein frame, the
Christoffel symbols are given by

\begin{equation}
\{^{i}_{tt} \} = - {1 \over 2} g^{i\rho}\partial^\rho h_{tt} =
{1 \over 2}\partial_\rho \ln(g_{tt}).
\end{equation}

\noindent
 Similarly, for the contortion the nonvanishing part is

\begin{equation}
K^{\hspace{.2 true cm}\rho}_{(tt)} = -\frac{\tilde \phi'}{2\tilde
\phi}\sim  \alpha \phi '=  \frac{1}{2}\sqrt{\frac{\lambda}{\pi}}
m\left[ \frac{\sqrt{\frac{\lambda}{\pi}}\rho \cot (\sqrt{\frac{\lambda}
{\pi}} \rho) -1]}{\sqrt{\frac{\lambda}{\pi}} m\rho +
\sin(\sqrt{\frac{\lambda}{\pi}} \rho)}\right]\; .
\end{equation}

We also note that the gravitational pull is related to the $h_{tt}$ component
that has an explicit dependence on the torsion, as shown in Eq.(\ref{htt}).
From the last equation, the acceleration that the SMSCS exerts on a test particle
can be explicitly written as

\begin{equation}
a_{ smscs} = -  m \sqrt{{\lambda \over \pi}}  \frac{ \sqrt{{\lambda
\over \pi}} \rho  \cot \sqrt{\frac{\lambda}{\pi}} \rho -1}
{\left[\sqrt{\frac{\lambda}{\pi}}m\rho + \sin \sqrt{\frac{\lambda}{\pi}}\rho
\right]} \; .
\label{force}
\end{equation}

\begin{figure}
\centering
\mbox{{\epsfig{figure=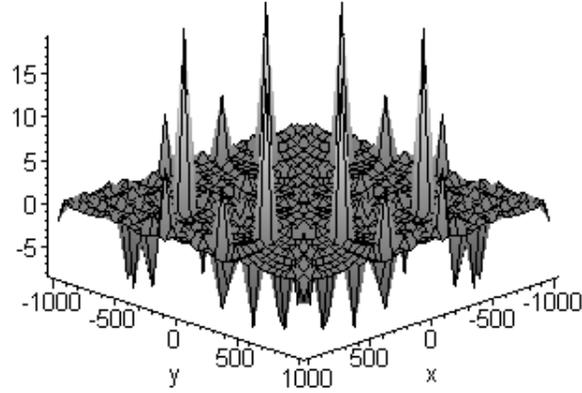,width=.8 \textwidth}}}
\vskip -1 cm
\caption{\label{fig3}  Acceleration as a function of x and y, for $ m=
0.2$,\, \,$\sqrt{\frac{\lambda}{\pi}}=0.2 $ and $\alpha = 0.5$, with x and y
in the range  \, \, $-1000 \leq x \leq 1000$ and  $-1000\leq y \leq 1000$.}
\end{figure}

\begin{figure}
\centering
\mbox{{\epsfig{figure=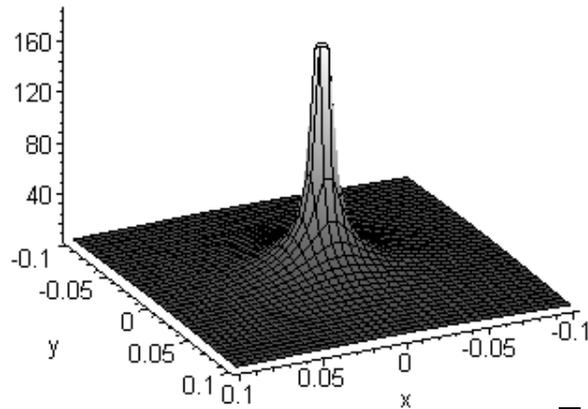,width=.8\textwidth}}}
\vskip -1 cm
\caption{\label{fig4}  Acceleration as a function of x and y for $ m=
0.2$,\, \, $\sqrt{\frac{\lambda}{\pi}}=0.2 $ and $\alpha = 0.5$, with x and
y in the range \, \, $-0.1 \leq x \leq 0.1$ and  $-0.1\leq y \leq 0.1$.}
\end{figure}

This solution is shown Fig.(\ref{fig3}), for a large range. We can also
consider the case which corresponds to a region where $ \sqrt{\frac{\lambda}
{\pi}} \rho$ is very small. In this case, the aceleration can be represented
by Fig.(\ref{fig4}). By comparing these two graphs one can see that the
perturbations introduced by the vortex are perceptible near the string.

A quick glance at the last equation allows us to understand the
essential r\^ole played by torsion  in the context of the present
formalism. If torsion is present, even in the case in which the
string has no current, an attractive gravitational force appears.
 In the context of the SMSCS, torsion acts in such a way so as
to enhance the force that a test particle experiences in the space-time
of a SMSCS. Assuming that this string survives inflation and is stable,
this enhancement may have meaningful astrophysical and cosmological
effects, as for instance, it may in principle influence the process
of structure formation in the universe.

\newpage

\section{Conclusion}

It is possible that torsion had had a physically relevant role during
the early stages of the Universe's evolution. Along these lines, torsion
fields may be potential sources of dynamical stresses which, when coupled
to other fundamental fields; for example, the gravitational and scalar fields,
might have an important role during the phase transitions leading to the
formation of topological defects such as the SMSCS here under consideration.
Therefore, it seems stimulating to investigate basic scenarios involving
production of topological defects within the context of scalar-tensor
theories of gravity with torsion.

 It is likely that ordinary cosmic strings could actively perturb the CMB.
But these strings cannot be wholly responsible for either the CMB temperature
fluctuations or the large scale structure formation of matter. Nonetheless,
in the case of supermassive screwed cosmic strings because of the fact that the
tension allowed in this situation is much larger than the one corresponding
to the ordinary cosmic, and that the torsion, as well as the scalar field, has
a non-negligible contribution to the dynamics around the string, one can expect
that such a SMSCS can induce perturbations in the universe matter and radiation
distribution which are larger than in the case of the ordinary cosmic string. In
spite of this, the perturbations by this way driven are certainly not enough to
leave significant imprints in the CMB anisotropies as implied by observations.

Thus, from the obtained results we conclude that the torsion as well as the
scalar field have a small but non-negligible contribution to the particle dynamics
as seen from its modification of the geodesic equation, which is obtained from the
contortion term and from the scalar field, respectively. From a physical point of
view, these contributions certainly are important and must be considered.

 The main motivation to consider this scenario is related to the fact that
scalar-tensor gravitational fields are important for a consistent description
of gravity, at least at sufficiently high energy scales. Further, the torsion
can either induce some physical effects, as the strongest acceleration of particles
moving around the string shown above, which could be important at some energy
scale, as for example, in the low-energy limit of string theory.

\section*{Acknowledgements}

C. N. Ferreira would like to thank (CNPq-Brasil) for financial support.
V. B. Bezerra would like to thank CNPq and FAPESQ-PB/CNPq(PRONEX) for
partial financial support.  H. J. Mosquera Cuesta thanks ICRA/CBPF/MCT
for the institutional support through the PCI/CNPq Fellowships Programme.

\end{document}